\documentclass[fleqn,usenatbib]{mnras}

\usepackage{newtxtext,newtxmath}

\usepackage[T1]{fontenc}

\DeclareRobustCommand{\VAN}[3]{#2}
\let\VANthebibliography\thebibliography
\def\thebibliography{\DeclareRobustCommand{\VAN}[3]{##3}\VANthebibliography}

\usepackage{graphicx}	
\usepackage{amsmath}	
\usepackage{multirow}
\usepackage{pdflscape}

\title[Testing mass-discrepancy problem of SESNe]{How to solve the mass-discrepancy problem of SESNe - I. Testing model approximations}

\author[A. P. Nagy ]{
Andrea P. Nagy,$^{1}$\thanks{E-mail: nagyandi@titan.physx.u-szeged.hu}
\\
$^{1}$Department of Experimental Physics, University of Szeged, Dom ter 9, Szeged, 6720, Hungary\\
}

\date{Accepted XXX. Received YYY; in original form ZZZ}

\pubyear{2022}

\begin{document}
\label{firstpage}
\pagerange{\pageref{firstpage}--\pageref{lastpage}}
\maketitle

\begin{abstract}
Here, we present a systematic study of 59 stripped-envelope supernovae (SESNe) (including Type IIb, Ib, Ic, and transitional events) to map a possible reason for the so-called mass-discrepancy problem. In this scenario, we assume the tension between the estimated ejected masses from early- and late-time light curves (LC) is due to  approximations generally used in analytical models. First, we examine the assumption that the R-band light curve is indeed a good approximation of the bolometric light curve. Next, we test the generally used assumption that rise-time to maximum brightness is equal to the effective diffusion time-scale that can be used to derive the ejecta mass from the early LC. In addition, we analyze the effect of gamma-ray and positron-leakage, which play an important role in forming the shape of the tails of SESNe, and also can be crucial to gaining the ejecta masses from the late-time LC data. Finally, we consider the effect of the different definitions of velocity that are needed for the ejecta mass calculations.
\end{abstract}




\section{Introduction}
It is generally believed that the death of massive stars (M > $8 M_\odot$) form core-collapse supernovae that can be divided into two main groups based on their spectral features. Those explosions that show hydrogen lines in their spectra are classified as Type II, hence those objects that are not labeled as Type Ib/Ic supernovae \citep[e.g.,][]{Filippenko, Gal-Yam}. 

Type Ib and Ic explosions can be separated by the presence or absence of helium in their spectra, however, there are a number of events that can be classified in both types (e.g. SN 1999ex, SN 2002ap, SN 2007gr). That fact indicates that the exact nature of Type Ib/Ic progenitors is not revealed yet. However, the lack of prominent hydrogen and/or helium indicates that during their evolution, the progenitor stars lost most of their outer layers due to interaction with a binary companion, strong stellar wind, or other severe mass-loss processes \citep[e.g.,][]{Podsiadlowski, Clocchiatti, Eldridge, Smartt}. Moreover, the light curve and spectral properties of Type Ib/Ic supernovae also imply relatively low ejected masses \citep[e.g.,][]{Dessart, Taddia, Lyman, Prentice}. Due to this latter nature Type Ib/Ic supernovae are also referred to as stripped-envelope SNe (SESNe) \citep{Clocchiatti}. Here, it should be noted that Type IIb supernovae, which are identified as a transitional group between hydrogen-rich and hydrogen-poor objects, can also be specified as SESNe, because of their low-mass hydrogen ejecta. Thus, in this paper, we use the terminology of stripped-envelope SNe for both Type Ib/Ic and Type IIb explosions.   

Even though the event rate of SESNe is relatively high (about $25 \%$) among core-collapse supernova  \citep[e.g.,][]{Shivvers}, we still do not understand some basic features of these objects. Theoretical studies \citep[e.g.,][]{Clocchiatti, Wheeler} revealed that there is a discrepancy in the derived ejected masses from early- and late-time light curve (LC) fits of stripped-envelope supernovae. Namely, estimations from the post-maximum light variation show systematically higher values than that can be deduced from the light curve peaks by the “Arnett’s rule” \citep[e.g.,][]{Khatami}. To solve this problem, we should take into account two different scenarios. First, it is plausible that the mass-discrepancy occurs due to the limitations and initial boundary conditions of the applied semi-analytic models (e.g., constant Thompson-scattering opacity, spherical symmetry). On the other hand, this mass-discrepancy may have a true physical cause, such as low-mass, low-density ejected, or circumstellar matter (CSM) around the supernova remnant or asymmetries in the explosion \citep[][]{Turatto, Sollerman}. Here we concentrate on the first plausible scenario related to the basic assumptions of the generally used semi-analytic light curve models  
to understand better the mass-discrepancy problem of stripped-envelope SNe from a theoretical point of view.

The paper is organized as follows: in Section 2 we present the data sample. In Section 3 we describe the analytical basis of our model, and explain how we obtain ejected masses from early- and late-time light curves. Section 4 presents our finding related to the difference between the commonly applied model assumptions and our modified initial conditions for both the early- and late-time light curve models. In Section 5 we compare the ejecta masses gained from different approximations. Finally, in Section 6 we summarize our conclusion. In addition, an Appendix is also included where we collect figures showing our best model fits of each supernova.

\section{Data Set}

\subsection{Sample Selection}
\label{sec:maths} 
We begin the data selection by assembling a “master list” of all transients
classified as Type Ib/Ic or Type IIb SLSNe in the Open Supernova Catalog \citep[][]{Guillochon}, regardless of whether they have been included in a refereed publication or not.  This first list of the available SESNe contained 903 events, but most were poorly sampled or missing early- or late-time photometric data. From this, we select the stripped-envelope supernovae with light curves that are well sampled around and after the peak to determine the ejected masses from both early- and late-time light curves. Thus, we identify 59 total SESNe with sufficient data listed in Table \ref{tab:1}. - \ref{tab:5}. Note that, during the analysis, we divided these supernovae into the following groups: Type Ic, Type Ib, Type Ibc, Type IIb and peculiar (Ic-BL and Ibc-pec). We also separate these supernovae according to their global light curve properties, similar objects, that show increasing or bumpy late-time bolometric light curves (group 1), and the ones that do not (group 2).   

\begin{table}
	\centering
	\caption{Basic properties of Type IIb SNe}
	\label{tab:1}
    \begin{tabular}{ lccc }
    \hline
    SN Name & Filter Coverage  & Distance (Mpc) & Group\\
    \hline
     SN1993J & UBVRIJHKL  & 3.31 &  1\\
     SN1996cb & BVR & 3.8 & 2 \\
     SN2004ex & BVJHYugri & 72.8 & 2 \\
     SN2006T & BVJHYugri & 32.9 & 2 \\
     SN2006el & BVRr'i' & 76.2 & 2 \\
     SN2008aq & UBVJHYugri & 26.0 & 2 \\
     SN2008ax & UBVRIJHugri & 5.81 & 2 \\
     SN2009mg & UBV & 36.4 & 2 \\
     SN2011dh & UBVRIJHK & 7.30 & 2 \\
     SN2011fu & UBVRI & 82.98 & 2 \\
     SN2011hs & UBVRI & 24.5 & 1 \\
     SN2012P & UBVgriz & 27.4 & 1 \\
     SN2013df & BVRIg'r'i'z'& 17.9 & 2 \\
     SN2016gkg & UBVRIgri & 26.4 & 2 \\ 
    \hline
\end{tabular}
\end{table}    

 \begin{table}
	\centering
	\caption{Basic properties of Type Ib SNe}
	\label{tab:2}
    \begin{tabular}{ lccc }
    \hline
    SN Name & Filter Coverage  & Distance (Mpc) & Group\\
    \hline
     SN1983N &  UBV  &  4.60 & 1\\
     SN1999dn &  UBVRIJHK  &  33.30 & 2\\
     SN2004gq &  UBVRJHugri  & 24.86 & 2\\
     SN2007C & UBVRJHugri   &  23.47 & 2\\
     SN2007uy & UBVJH   &  28.98 & 2\\
     SN2008D &  UBVRIJH  &  29.70 & 2\\
     SN2009jf &  BVJHri  &  34.44 & 2\\
     iPTF13bvn &  UBVRIgri  &  19.94 & 2\\
    \hline
\end{tabular}
\end{table}    

 \begin{table}
	\centering
	\caption{Basic properties of Type Ibc SNe}
	\label{tab:3}
    \begin{tabular}{ lccc }
    \hline
    SN Name & Filter Coverage  & Distance (Mpc) & Group\\
    \hline   
     SN1962L & UBV & 5.83 & 2\\
     SN1998bw & UBVRI & 37.9 & 1\\
     SN1999ex & UBVRI &  43.3& 2\\
     SN2002ap & UBVRI & 6.7 & 2\\
     SN2005hg & UBVRJHK & 81.0 & 1\\
     SN2006ep & BVJHugri & 58.9 & 2\\
     SN2007Y & UBVugri & 19.38 & 2\\
     SN2007gr & UBVRIJH & 9.3 & 2\\
     SN2009iz & UBVJHri & 63.5 & 2\\
     DES17C1ffz & griz & 427.5 & 2\\
    \hline
\end{tabular}
\end{table}    

 \begin{table}
	\centering
	\caption{Basic properties of Type Ic SNe}
	\label{tab:4}
    \begin{tabular}{ lccc }    
    \hline
    SN Name & Filter Coverage  & Distance (Mpc) & Group\\
    \hline
     SN1994I & UBVRI & 7.8 & 2 \\
     SN2004aw & UBVRI & 74.3 & 2 \\
     SN2004dn & BVRI & 50.7 & 2 \\
     SN2004fe & BVRgri  &  76.13& 2\\
     SN2004ff & BVRgri & 72.5  & 2\\
     SN2006lc & BVg'r'i'z' & 59.2 & 2\\
     SN2007cl & Vr'i' & 98.7 & 2\\
     SN2009bb & BVRIJHugri & 40.2 & 1\\
     SN2010bh &  BVRiz & 274.1 & 1\\
     SN2010gx & u'g'r'i'z' & 1181.3 & 2\\
     SN2010md & Bgriz & 468.8 & 2\\
     SN2011bm & UBRIugriz & 99.0 & 1\\
     SN2011ke & UBVugriz & 697.8 & 1\\
     SN2011kg & UBJugriz & 968.7 & 1\\
     SN2013dg & griz & 1389.0 & 1\\
     SN2013ge & UBVRIri & 15.95 & 2\\
     SN2014L & UBVRI & 14.4 & 1\\
     SN2014ad & UBVRI & 25.3 & 2\\
     SN2015bn & UBVRIJHKu'g'r'i'z' & 544.8 & 1\\
     PTF15dtg & gri & 241.1 & 1\\
     SN2017ein & UBVRIgr & 12.69 & 2\\
    \hline
\end{tabular}
\end{table}    

 \begin{table}
	\centering
	\caption{Basic properties of peculiar SESNe}
	\label{tab:5}
    \begin{tabular}{ lccc }    
    \hline
    SN Name & Filter Coverage  & Distance (Mpc) & Group\\
    \hline  
     SN1997ef & V & 53.5 & 2\\
     SN2003dh & UBVRIJH & 836.0 & 2\\
     SN2007bi & BVRI & 530.0 & 1\\
     SN2015U & BVRI &60.3 & 2\\
     SN2016coi & UBVRIugriz & 15.8 & 1\\
     SN2019dge & ugriz & 93.0 & 2\\
    \hline 
\end{tabular}
\end{table}

\subsection{Quasi-bolometric Light Curve}
To build the quasi-bolometric light curves (LC) of the selected 59 supernovae, we applied the following method. First, we collected the measured magnitudes in all available photometric bands. Then, these data were converted into fluxes using extinctions, distances, and zero points based on \cite[][]{Bessell}. The extinction values and distances in each case were taken from NASA/IPAC Extragalactic Database (NED). Photometric data were not equally sampled across all filters, thus, at epochs when the measurements were missing, we linearly interpolated the flux using nearby data. The trapezoidal rule was applied while integrating over all wavelength with the assumption that flux reaches zero at 2000 $\mathring{A}$. The infrared contribution was taken into account by the exact integration of the Rayleigh-Jeans tail from the wavelength of the last available photometric band to infinity.

\section{Model Basis}
\label{sec:model}
In this study we used our self-developed semi-analytic LC code \citep{Nagy, Nagy1}, which is based on the radiative diffusion model originally developed by \cite{Arnett}. During the light curve fits, we assumed homologous expansion, spherically symmetric supernova ejecta, constant density profile, and constant opacity. We also suppose that the bolometric luminosity ($L_{bol}$) of the stripped-envelope supernovae is only determined by the energy generated from the radioactive heating of nickel and cobalt. Hence, the spin-down of a low-energy magnetar could be an alternative energy source \citep[e.g.,][]{Kasen, Inserra} beside radioactivity (see details in our forthcoming paper). 

Because of the supposedly low ejected mass, the effect of gamma-ray and positron leakage was also taken into account as 
\begin{align}
        &L_{bol} = L_{Ni}\ (1 - exp(-T_0^2/t^2))\ +  \\ \nonumber
        &+  M_{Ni}\ \left[E_{ki} + E_{Co}\ (1 - exp(-T_0^2/t^2))\right](1 - exp(-T_p^2/t^2))\ ,
\end{align}
where $L_{Ni}$ is energy released from radioactive heating \citep[][]{Arnett, Nagy}, $E_{ki}$ and $E_{Co}$ are constants \citep[][]{Nadyozhin}, and $T_g$ and $T_p$ factors are referred to the characteristic time-scale of gamma-ray and positron leakage, respectively. 

Using this fitting method, we should be able to obtain the approximate mass of the supernova ejecta, but for most of SESNe, this solution is quite complicated due to their peculiar luminosity gradient at late-times. Hence in this context, we tried not only to simulate the entire bolometric light curve as accurately as possible, but we also adjusted the ejected masses separately from fits to the peak and tail data. 

To estimate ejected masses from the early-time LC, we applied the following equation:
\begin{equation}
    M_{ej} \approx \frac{\beta\ c}{2\ \kappa}\ v_{sc}\ t_d^2\ ,
\end{equation}
where $\beta = 13.8$ is an integration constant, $c$ is the speed of light, $\kappa$ is the constant opacity, $v_{sc}$ is the scaling velocity and $t_d$ is the effective diffusion time-scale. In the literature the accepted definition of $v_{sc}$ for a constant density ejecta is
\begin{equation}
    v_{sc}^2 = \frac{10\ E_{kin}}{3\ M_{ej}}\ ,
\end{equation}
where $E_{kin}$ is the kinetic energy of the supernova explosion. However, the scaling velocity is not related directly to any observable velocities, thus, during the calculations, other definitions were also used (see details in Sect. \ref{sec:velocity}).   

For the late-time tail, we adopt the following equation to calculate the ejected mass:
\begin{equation}
    M_{ej} \approx \sqrt{\frac{T_0^2\ E_{kin}}{C\ \kappa_\gamma}} = \frac{4\ \pi\ T_0^2\ v_{sc}^2}{3\ \kappa_\gamma}\ ,
\end{equation}
where $\kappa_\gamma = 0.028\ cm^2/g $ is the gamma-ray opacity \citep{Colgate}, and $C$ is a dimensionless constant dependent on the slope of the density profile. For example $C \approx 0.07$ for a constant density ejecta.

\section{Testing usual model assumptions}
In order to map the main reasons for the tension between the estimated ejected masses from early- and late-time light curves, we tested separately assumptions generally used for both. 

\subsection{R-band vs. Bolometric LC}
According to some preceding papers \citep[e.g.,][]{Wheeler}, the R-band light curve is a good approximation of the bolometric light curve. To test the effect of this assumption, we compare the bolometric light curves of 25 SNe with their R-band luminosity data. 

To gain R-band luminosities, we first calculate the absolute magnitudes ($M$) using the distance modulus. Than we take into account the bolometric correction as $BC \approx -0.3^m$ \citep[][]{Lyman14} to determine the bolometric magnitude of the supernova ($M_{bol} = M + BC$). And finally, the following equation was used to calculate the luminosities from R-band data:      
\begin{equation}
    \log L = -\frac{M_{bol} - M_\odot}{2.5}\ + \log L_\odot\ ,
\end{equation}
where $M_\odot$ and $L_\odot$ are the bolometric magnitude and the luminosity of the Sun, respectively. 

 \begin{figure}
	\includegraphics[width=\columnwidth]{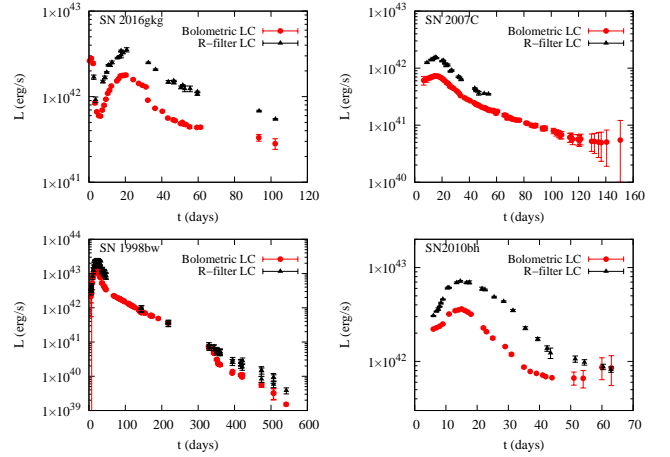}
    \caption{Example for the comparison of the R-band (black dots) and bolometric (red dots) LCs of different supernova types. }
    \label{fig:R_band}
\end{figure}

For this analysis, we compare the maximal luminosities and the widths of the peak for both the R-band ($\Delta t_R$) and bolometric ($\Delta t_{bol}$) light curves. As can be seen in Table \ref{tab:6}, we get a slightly different result for the different supernova types. For Type IIb and Ibc supernovae, the rise times, the width of the peaks, and the slope of the late-time LCs are not correlated (see e.g. in Fig. \ref{fig:R_band}.) in most cases. However, for Type Ib and Ic, only the near maximum light curves may cover each other by calibrating the luminosities. Thus, scaling the R-band light curve is not usually a plausible solution.
 
  \begin{table}
	\centering
	\caption{Comparing R-band and bolometric light curve}
	\label{tab:6}
    \begin{tabular}{ lccc }    
    \hline
    SN Name & $\Delta t_R$ (days) & $\Delta t_{bol}$ (days) & $L_R/L_{bol}$\\
    \hline  
     SN1996cb & 27.7 & 23.3 & 1.83\\
     SN2006el & 38.4 & 25.9 & 1.91\\
     SN2011hs & 19.3 & 19.3 & 2.81\\
     SN2016gkg & 35.5 & 28.6 & 1.89\\
    \hline
     SN1998bw & 23.1 & 18.9 & 1.67\\
     SN1999ex & 28.4 & 24.3 & 2.15\\
     SN2002ap & 23.2 & 22.8 & 2.27\\
     SN2005hg & 24.5 & 24.5 & 2.07\\
     SN2007gr & 20.9 & 20.9 & 5.58\\
    \hline
     SN1999dn & 21.0 & 21.0 & 2.02\\
     SN2007C & 17.2 & 17.2 & 2.04\\
     SN2008D & 26.1 & 26.1 & 2.39\\
     \hline
     SN1994I & 17.1 & 17.1 & 2.32\\
     SN2004aw & 27.3 & 27.3 & 2.33\\
     SN2004dn & 15.7 & 15.7 & 2.57\\
     SN2004fe & 14.9 & 12.9 & 2.39\\
     SN2004ff & 9.8 & 9.8 & 2.55\\
     SN2009bb & 20.7 & 20.7 & 1.77\\
     SN2010bh & 19.7 & 13.9 & 2.07\\
     SN2011bm & 50.7 & 48.5 & 1.74\\
     SN2013ge & 30.3 & 30.3 & 2.34\\
     SN2014L & 22.4 & 22.4 & 2.53\\
     SN2014ad & 14.8 & 14.8 & 2.79\\
     SN2016coi & 20.1 & 20.1 & 2.95\\
     SN2017ein & 21.4 & 21.4 & 2.26\\
     \hline
\end{tabular}
\end{table}
 
For quantitative analysis, we fit both the R-band and bolometric data and compare the ejected masses, which indicates a significant (up to 30$\%$) difference in the derived results. However, the ratio between $M_{ej}$ gained from early- and late-time fits show no notable differences. Thus, this approximation is not fully adequate to estimate the physical properties of the individual supernovae, but it does not make a considerable effect on solving the mass-discrepancy problem. 

\subsection{Velocity Assumptions}
\label{sec:velocity}

As it was mentioned above, to calculate ejected masses from both early- and late-time light curves, we need to know the value of the scaling velocity that is not related directly to any observable properties. Thus, during the calculations an approximation for $v_{sc}$ should be taken into account. 

One possibility is that the scaling velocities are derived from LC modeling. However, these values are always somewhat uncertain because of the parameter correlations \citep[][]{Nagy} that may lead to a significant systematic error in the calculated ejected masses. Here, we create generally acceptable fits for the entire supernova light curves, and its $v_{sc}$ was used as a reference. 

Another plausible solution could be that the photospheric velocity ($v_{ph}$) is approximately equal to the scaling velocity. However, in practice, $v_{ph}$ could be diverse for different chemical elements, and it can also be affected by the density distribution as well as the opacity. Hence, this assumption does not describe well the fixed scaling velocity.

A further assumption could be that we use an average expansion velocity ($\overline{v}$), depending on the supernova type, as the scaling velocity. For example for Type Ib/IIb, Type Ibc and Type Ic $9000 $ km/s, $9500 $ km/s  and $10000 $ km/s, respectively. This approximation could define the velocities of such supernovae that do not have spectroscopic data.

Nevertheless, these definitions of the velocity show significantly different values for the individual supernovae (see in Table \ref{tab:vel}). Hence the ejected mass is sensitive to the choice of the velocity, we calculate $M_{ej}$ with both approximations, and compare all of the derived results for both early- and late-time models.

\begin{table}
	\centering
	\caption{Velocity values of SESNe}
	\label{tab:vel}
    \begin{tabular}{ lccc }
    \hline
    SN Name & $\overline{v}$ (km/s) & $v_{ph}$ (km/s)& $v_{sc}$ (km/s)\\
    \hline
     SN1993J & 9000  & 7500 & 14100\\
     SN1996cb & 9000 & 9000 & 10900 \\
     SN2004ex & 9000 & - & 14100 \\
     SN2006T & 9000 & 7500 &  17500 \\
     SN2006el & 9000 & 11000 & 12900 \\
     SN2008aq & 9000 & - & 13700 \\
     SN2008ax & 9000 & 8000 & 25100 \\
     SN2009mg & 9000 & 10000 & 16000 \\
     SN2011dh & 9000 & 6900 & 21800 \\
     SN2011fu & 9000 & 11000 & 16400 \\
     SN2011hs & 9000 & 8000 & 14100 \\
     SN2012P & 9000 & - & 16200 \\
     SN2013df & 9000 & 10800 & 13800 \\
     SN2016gkg & 9000 & - & 12000 \\ 
    \hline 
     SN1983N &  9000  &  7800 & 10400\\
     SN1999dn &  9000  &  10000 & 11600\\
     SN2004gq &  9000  & 13000 & 7000\\
     SN2007C & 9000   &  11000 & 14500\\
     SN2007uy & 9000  &  14000 & 4700\\
     SN2008D &  9000  &  10000 & 8600\\
     SN2009jf &  9000  &  10000 & 5900\\
     iPTF13bvn &  9000  &  8700 & 7500\\
    \hline
     SN1962L & 9500 & 6200 & 10200\\
     SN1998bw & 9500 & 14000 & 8200\\
     SN1999ex & 9500 & - & 12500\\
     SN2002ap & 9500 & 15000 & 9700\\
     SN2005hg & 9500 & 9000 & 9500\\
     SN2006ep & 9500 & 9500 & 8900\\
     SN2007Y & 9500 & 9000 & 12400\\
     SN2007gr & 9500 & 6400 & 6000\\
     SN2009iz & 9500 & - & 7000\\
     DES17C1ffz & 9500 & - & 11500\\
    \hline 
     SN1994I & 10000 & 11500 & 8100 \\
     SN2004aw & 10000 & 16000 & 8500 \\
     SN2004dn & 10000 & 12500 & 12000 \\
     SN2004fe & 10000  & 11000 & 14300\\
     SN2004ff & 10000 & 11000  & 8500\\
     SN2006lc & 10000 & - & 12500\\
     SN2007cl & 10000 & - & 10000\\
     SN2009bb & 10000 & 18000 & 10900\\
     SN2010bh &  10000 & 30000 & 11600\\
     SN2010gx & 10000 & 22900 & 8000\\
     SN2010md & 10000 & - & 7800\\
     SN2011bm & 10000 & 7500 & 9000\\
     SN2011ke & 10000 & - & 7500\\
     SN2011kg & 10000 & - & 8000\\
     SN2013dg & 10000 & 18000 & 8500\\
     SN2013ge & 10000 & 10000 & 12000\\
     SN2014L & 10000 & 7600 & 14700\\
     SN2014ad & 10000 & - & 13700\\
     SN2015bn & 10000 & 6500 & 9800\\
     PTF15dtg & 10000 & 6000 & 10800\\
     SN2017ein & 10000 & 9300 & 14900\\
     \hline
     SN1997ef & 10000 & 9500 & 9200\\
     SN2003dh & 10000 & - & 11100\\
     SN2007bi & 10000 & - & 5000\\
     SN2015U & 9000 & - & 18600\\
     SN2016coi & 10000 & 14000 & 6500\\
     SN2019dge & 9000 & 8000 & 25400\\
    \hline 
\end{tabular}
\end{table}

\subsection{Gamma-ray vs. Positron-leakage}
 It is a generally used approach in the literature that we neglect the partial leakage of the positrons, which is a good approximation as far as the mass of the ejecta is high enough (e.g., for Type IIP) to forbid the fast leakage of the gamma-rays. But for SNe with lower ejected masses, like stripped-envelope ones, the contribution of the positrons will be significant.

\begin{figure}
	\includegraphics[width=\columnwidth]{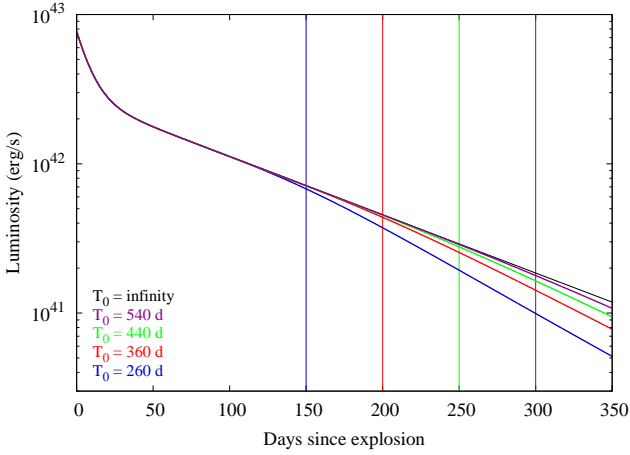}
    \caption{Examples for LC tail deflections due to different $T_0$ values. Vertical lines show the time-scale, where a particular colored line starts to differ from the Ni-Co tail. The $T_0 = \infty$ (black curve) represents the lack of the gamma- and positron-leakage. Thus, this curve is equal to the nickel-cobalt tail.}
    \label{fig:tail}
\end{figure}

 The motive, why this approximation seems reasonable is that the time-scale of the positron leakage is much longer than the average length of the SESNe light curves. Thus, positrons should not play an important role in forming the shape of the late-time light curves because of their enormous (around 400 - 2000 days) $\ T_p$ values.  However, from a modeling point of view using both gamma-ray and positron leakage does not just add an extra energy source at very late times, but also reduces $T_0$, which not just causes lower derived ejected masses from the tail, but also reduces the time-scale, when the late-time light curve diverges from the nickel-cobalt tail (Fig. \ref{fig:tail}.).  

To test this theory, we fit the late-time light curve of all SESNe with a $\chi^2$ method using only gamma-ray, as well as both gamma-ray and positron leakage, and then calculate the ejected masses from them. For simplicity we distinguish the most important parameters ($T_0$, $M_{Ni}$ and $M_{ej}$) for the case, where only gamma-ray leakage is employed (and the effect of positrons are just ignored) as $T_\gamma$, $M_{Ni,\gamma}$ and $M_\gamma$, while we use $T_{+}$, $M_{Ni,+}$ and $M_{+}$ for the other scenario. 

As can be seen in the Appendix (Fig. \ref{fig:lc1} - \ref{fig:lc5}), both scenario are capable of reproducing the same late-time LCs. However, using both processes causes significantly lower nickel masses (see in Table \ref{tab:gamma}) that is more reasonable for SESNe \citep[e.g.,][]{Sawada}. It is also noticeable that some fits needed an extremely high nickel mass to generate the LC which may suggest an extra energy input besides radioactive decay. The detailed analysis of this phenomenon is beyond the scope of this paper (see details in our forthcoming paper). 

The results also show that the addition of the positron-leakage reduces the $T_0$ value in most cases, except for two objects (SN 2013df, SN 2019dge). For these supernovae, inconsistency may occur due to the lack of observational data at later times. Moreover, there are some supernovae that despite their low ejecta masses, follow the Ni-Co tail. Thus, their late-time fits result in the upper limit (600 days) for $\ T_0$, which indicates that basically all gamma-rays are trapped inside the ejecta.

\begin{table}
	\centering
	\caption{$T_0$ values from different gamma-ray leakage assumptions}
	\label{tab:gamma}
    \begin{tabular}{ lcccc }
    \hline
    SN Name & $T_{\gamma}$ (d)& $T_{+}$ (d) & $M_{Ni,\gamma}$ ($M_\odot$) & $M_{Ni,+}$ ($M_\odot$)\\
    \hline
     SN1993J & 114.7  & 113.6 & 0.085 & 0.042\\
     SN1996cb & 109.9 & 107.1 & 0.0034 & 0.0017\\
     SN2004ex & 96.0 & 91.2 & 0.063 & 0.032\\
     SN2006T & 101.6 & 99.1 & 0.06 & 0.03\\
     SN2006el & 117.5 & 117.4 & 0.053 & 0.026\\
     SN2008aq & 96.0 & 93.8 & 0.03 & 0.015\\
     SN2008ax & 96.3 & 92.7 & 0.089 & 0.045\\
     SN2009mg & 121.6 & 115.8 & 0.026 & 0.013\\
     SN2011dh & 112.5 & 109.2 & 0.048 & 0.024\\
     SN2011fu & 191.0 & 190.6 & 0.2 & 0.098\\
     SN2011hs & 600.0 & 600.0 & 0.008 & 0.008\\
     SN2012P & 138.6 & 120.0 & 0.027 & 0.014\\
     SN2013df & 49.9 & 66.7 & 0.05 & 0.02\\
     SN2016gkg & 203.3 & 202.9 & 0.055 & 0.027\\ 
    \hline
     SN1983N &  149.1  &  131.1 & 0.29 & 0.16\\
     SN1999dn &  250.0  & 250.0 & 0.024 & 0.012\\
     SN2004gq &  127.2 & 114.0 & 0.029 & 0.015\\
     SN2007C & 113.8 &  108.5 & 0.019 & 0.0096\\
     SN2007uy & 157.9 & 154.9 & 0.0133 & 0.0066\\
     SN2008D &  95.9  & 95.7 & 0.0158 & 0.008\\
     SN2009jf &  141.0  & 138.8 & 0.2 & 0.099\\
     iPTF13bvn &  135.9  & 135.4 & 0.017 & 0.009\\
    \hline
     SN1962L & 219.9 & 43.7 & 0.0029 & 0.0019\\
     SN1998bw & 156.6 & 151.8 & 0.3 & 0.15\\
     SN1999ex & 49.4 & 42.0 & 0.042 & 0.025\\
     SN2002ap & 125.6 & 125.4 & 0.043 & 0.022\\
     SN2005hg & 600.0 & 600.0 & 0.047 & 0.047 \\
     SN2006ep & 133.4 & 129.4 & 0.026 & 0.013\\
     SN2007Y & 80.5 & 66.4 & 0.026 & 0.015\\
     SN2007gr & 135.9 & 131.6 & 0.0102 & 0.0055\\
     SN2009iz & 111.0 & 110.8 & 0.114 & 0.057\\
     DES17C1ffz & 112.1 & 111.4 & 0.159 & 0.081\\
    \hline
     SN1994I & 92.8 & 88.9 & 0.0095 & 0.0048\\
     SN2004aw & 361.4 & 350.4 & 0.058 & 0.029\\
     SN2004dn & 107.3 & 103.7 & 0.026 & 0.013\\
     SN2004fe & 45.6  & 41.0 & 0.049 & 0.031\\
     SN2004ff & 600.0 & 600.0 & 0.0067 & 0.0067\\
     SN2006lc & 125.6 & 114.2 & 0.02 & 0.01\\
     SN2007cl & 72.8 & 70.3 & 0.057 & 0.029\\
     SN2009bb & 58.2 & 54.1 & 0.043 & 0.023\\
     SN2010bh &  600.0 & 600.0 & 0.046 & 0.046\\
     SN2010gx & 38.0 & 28.4 & 9.9 & 7.6\\
     SN2010md & 183.2 & 146.7 & 1.8 & 1.0\\
     SN2011bm & 99.0 & 89.7 & 0.57 & 0.2\\
     SN2011ke & 44.2 & 40.8 & 6.9 & 3.6\\
     SN2011kg & 600.0 & 600.0 & 1.8 & 1.8\\
     SN2013dg & 141.7 & 85.2 & 2.9 & 2.9\\
     SN2013ge & 185.1 & 184.8 & 0.0108 & 0.0054\\
     SN2014L & 600.0 & 600.0 & 0.0061 & 0.0061\\
     SN2014ad & 95.8 & 93.6 & 0.054 & 0.027\\
     SN2015bn & 600.0 & 275.9 & 14.8 & 10.8\\
     PTF15dtg & 600.0 & 600.0 & 0.2 & 0.2\\
     SN2017ein & 600.0 & 600.0 & 0.002 & 0.002\\
    \hline
     SN1997ef & 110.1 & 107.3 & 0.082 & 0.041\\
     SN2003dh & 330.0 & 219.5 & 0.4 & 0.2\\
     SN2007bi & 337.8 & 337.1 & 4.9 & 1.95\\
     SN2015U & 56.9 & 49.9 & 0.021 & 0.012\\
     SN2016coi & 157.9 & 152.8 & 0.072 & 0.036\\
     SN2019dge & 79.7 & 80.9 & 0.0094 & 0.0049\\
    \hline 
\end{tabular}
\end{table}

\subsection{Rise-time vs. Effective Diffusion Time-scale}
\label{sec:peak}
As it was mentioned in Section \ref{sec:model}., we fit the entire bolometric light curve of each supernova. As a fitting parameter, we derive the ejecta masses ($M_{ej}$) from which we were able to calculate the effective diffusion time-scales ($t_d$). So, hereafter we use these $t_d$ values as a reference to examine the contradiction between the effective diffusion time-scale and the rise time.

From a modeling point of view, we may neglect light curve fitting to determine the ejecta mass if we can somehow estimate the effective diffusion time-scale. An easy way to do so is if $t_d$ is assumed to be equal to the rise time to maximum brightness ($t_{peak}$). However, defining $t_{peak}$ value is quite uncertain due to the mostly indeterminable explosion time.

\begin{figure}
	\includegraphics[width=\columnwidth]{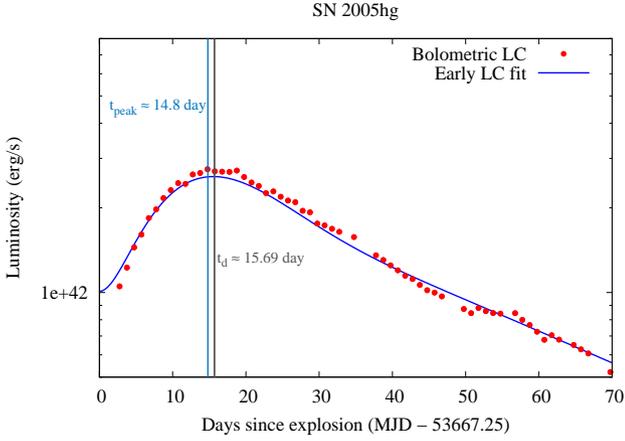}
    \caption{Difference between the rise-time and the effective diffusion time-scale for SN 2005hg, where we can estimate the explosion time quite certain because of a non-detection event.}
    \label{fig:dif}
\end{figure}

Moreover, if we assume that the rise time is roughly equal to $t_d$ we underestimate the mass. Because, in reality, the rise time is usually much shorter than the effective diffusion time-scale (Fig. \ref{fig:dif}), which results in lower masses as the expansion velocity is the same. As it can be seen in Table \ref{tab:t} sometimes (mainly for Type IIb SNe) an opposite relation ($t_{peak} > t_d$) occurs. This phenomenon may be explained by an unusually steep peak that is hard to fit properly, or a neglected second component (see details in \cite{Nagy1}) that prolongs the rise-time.

\begin{table}
	\centering
	\caption{Rise-time and effective diffusion time-scale of SESNe}
	\label{tab:t}
    \begin{tabular}{ lcc }
    \hline
    SN Name & $t_{d}$ (d)& $t_{peak}$ (d)\\
    \hline
     SN1993J & 19.61  & 20.6\\
     SN1996cb & 25.73 & 16.6 \\
     SN2004ex & 16.89 & 16.7 \\
     SN2006T & 18.44 & 14.0 \\
     SN2006el & 19.48 & 12.6 \\
     SN2008aq & 15.68 & 10.9 \\
     SN2008ax & 14.12 & 16.6 \\
     SN2009mg & 17.53 & 14.9 \\
     SN2011dh & 16.90 & 19.9 \\
     SN2011fu & 13.77 & 16.3 \\
     SN2011hs & 18.25 & 13.1 \\
     SN2012P & 12.15 & 10.5 \\
     SN2013df & 21.06 & 20.1 \\
     SN2016gkg & 20.65 & 17.6 \\ 
    \hline 
     SN1983N &  15.63  &  15.4\\
     SN1999dn &  19.77  &  14.0\\
     SN2004gq &  15.83  & 8.3\\
     SN2007C & 14.31   &  9.8\\
     SN2007uy & 18.57 & 15.7\\
     SN2008D &  18.92  & 16.1\\
     SN2009jf &  22.92  & 22.6\\
     iPTF13bvn &  15.79  & 14.1\\
    \hline
     SN1962L & 27.12 & 17.1\\
     SN1998bw & 14.89 & 14.0\\
     SN1999ex & 19.29 & 15.1\\
     SN2002ap & 16.51 & 10.1\\
     SN2005hg & 15.69 & 14.8\\
     SN2006ep & 16.94 & 12.2\\
     SN2007Y & 17.51 & 15.4\\
     SN2007gr & 17.48 & 11.8\\
     SN2009iz & 23.02 & 17.2\\
     DES17C1ffz & 20.43 & 23.4\\
    \hline
     SN1994I & 13.36 & 8.1 \\
     SN2004aw & 17.23 & 16.0 \\
     SN2004dn & 14.72 & 12.8 \\
     SN2004fe & 9.13  & 8.8\\
     SN2004ff & 19.61 & 10.6\\
     SN2006lc & 21.28 & 13.6\\
     SN2007cl & 21.49 & 14.4\\
     SN2009bb & 13.89 & 11.2\\
     SN2010bh &  15.27 & 13.0\\
     SN2010gx & 48.09 & 14.7\\
     SN2010md & 32.66 & 32.1\\
     SN2011bm & 27.52 & 27.1\\
     SN2011ke & 53.09 & 24.3\\
     SN2011kg & 43.61 & 23.1\\
     SN2013dg & 43.62 & 5.6\\
     SN2013ge & 25.92 & 16.5\\
     SN2014L & 19.90 & 11.9\\
     SN2014ad & 22.18 & 11.5\\
     SN2015bn & 85.37 & 60.2\\
     PTF15dtg & 22.47 & 21.9\\
     SN2017ein & 15.77 & 14.0\\
    \hline
     SN1997ef & 23.17 & 16.7\\
     SN2003dh & 8.37 & 2.1\\
     SN2007bi & 52.45 & 22.0\\
     SN2015U & 9.49 & 7.9\\
     SN2016coi & 23.99 & 16.7\\
     SN2019dge & 4.22 & 3.5\\
    \hline 
\end{tabular}
\end{table}    

\section{Comparing ejecta masses}
Using the results of the early- and late-time light curve calculations allows us to determine the ejecta masses of each scenario. During this process, different velocity estimations were used to compare as many model assumptions as possible. For early LC calculations, we adopt another crucial initial condition, namely, we assume that the average opacity is equal to $\kappa$ values of the entire light curve fits. However, the $M_{peak}$ also can be determined by an average Thomson-scattering or a mean hydrodynamic opacities \citep{Nagy2} for each supernova type, which definitely changes its value  (see details in our forthcoming paper).  

As can be seen in Table \ref{tab:mass1} - \ref{tab:mass3}, using the rise time approximation in most cases underestimates the ejecta masses, while both late-time methods usually overestimate $M_{ej}$. But, it is also clear, that applying gamma-ray and positron leakage somewhat reduces the difference between the derived ejected masses and gave more realistic nickel masses. 


\begin{table}
\centering
\caption{Ejected masses of SESNe calculated from the average expansion velocities ($\overline{v}$)}
\label{tab:mass1}
\begin{tabular}{lcccc}
\hline
  SN Name & $M_{ej}$ ($M_\odot$)  & $M_{peak}$ ($M_\odot$) & $M_\gamma$ ($M_\odot$) & $M_{+}$ ($M_\odot$) \\
 \hline
     SN1993J & 2.1 & 1.48  & 5.98  & 5.87 \\
     SN1996cb & 2.8 & 1.48 & 5.49 & 5.22 \\
     SN2004ex & 2.6 & 1.62 & 4.19 & 3.78 \\
     SN2006T & 2.3 & 1.37  & 4.69 & 4.47 \\
     SN2006el & 3.8 & 1.11 & 6.28 & 6.27 \\
     SN2008aq & 2.6 & 0.83 & 4.19 & 4.00 \\
     SN2008ax & 1.77 & 0.88 & 4.22 & 3.91\\
     SN2009mg & 1.9 & 1.55 & 6.72 & 6.10 \\
     SN2011dh & 1.4 & 0.92 & 5.76 & 5.42\\
     SN2011fu & 1.0 & 0.77 &  16.68 & 16.52 \\
     SN2011hs & 2.6 & 1.20 & -- & --\\
     SN2012P & 1.85 & 0.71 & 8.74  & 6.55  \\
     SN2013df & 2.37 & 1.41 & 1.13  & 2.02 \\
     SN2016gkg & 3.6 & 1.97 & 18.80 & 18.72\\ 
    \hline 
     SN1983N & 2.8 & 2.37 & 10.11 & 7.82 \\
     SN1999dn & 3.5 & 1.37 & 28.42 & 28.42 \\
     SN2004gq & 3.4 & 1.19  & 7.36 & 5.91 \\
     SN2007C & 2.3 & 0.67  & 5.89 & 5.35 \\
     SN2007uy & 6.2 & 8.63  & 11.34  & 10.91 \\
     SN2008D & 3.4 & 2.60 & 4.18 & 4.17 \\
     SN2009jf & 4.8 & 7.12  & 9.04 & 8.76 \\
     PTF13bvn & 4.8 & 4.62 & 8.40  & 8.34 \\
    \hline
     SN1962L & 2.9 & 1.07 & 27.15 & 1.07 \\
     SN1998bw & 4.7 & 4.84 & 12.43 & 11.68\\
     SN1999ex & 3.0 & 1.40 & 1.24 & 0.89\\
     SN2002ap & 6.8 & 2.48  & 7.99 & 7.97 \\
     SN2005hg & 3.0  & 2.70 & -- & --\\
     SN2006ep & 3.8 & 2.11  & 9.02 & 8.48 \\
     SN2007Y & 2.94 & 1.75 & 3.28 & 2.23\\
     SN2007gr & 4.7 & 3.43  & 9.36 & 8.78 \\
     SN2009iz & 4.1 & 3.12  & 6.24 & 6.22 \\
     DES17C1ffz & 3.7 & 4.04 & 6.37 & 6.29\\
    \hline
     SN1994I & 2.8 & 1.27 & 4.84 & 4.44\\
     SN2004aw & 6.5 & 6.63  & 73.33 & 68.82\\
     SN2004dn & 2.0 & 1.27 & 6.46 & 6.04 \\
     SN2004fe & 2.3 & 1.50 & 1.17 & 0.94\\
     SN2004ff & 2.8  & 0.97 & -- & -- \\
     SN2006lc & 2.91 & 0.96 & 8.86 & 7.32\\
     SN2007cl & 2.99 & 1.34  & 2.98 & 2.77\\
     SN2009bb & 2.7 & 1.62  & 1.90 & 1.64 \\
     SN2010bh & 2.1  & 1.31  & -- & --\\
     SN2010gx & 4.75 & 0.56 & 0.81 & 0.45\\
     SN2010md & 6.4 & 8.00 & 18.84 & 12.08 \\
     SN2011bm & 6.6 & 7.13 & 5.50 & 4.52\\
     SN2011ke & 5.42 & 1.53 & 1.10 & 0.93\\
     SN2011kg & 4.5  & 1.59 & -- &--\\
     SN2013dg & 3.92 & 0.08 & 11.27 & 4.08 \\
     SN2013ge & 3.2 & 1.06 & 19.24 & 19.17\\
     SN2014L & 2.25  & 0.55 & -- & --\\
     SN2014ad & 2.6 & 0.51 & 5.15  & 4.92 \\
     SN2015bn & 13.8 & 7.03  & -- & 42.74 \\
     iPTF15dtg & 4.2 & 3.72 & -- & --\\
     SN2017ein & 2.2 & 1.17 & -- & --\\
    \hline
     SN1997ef & 3.8 & 2.17  & 6.81 & 6.46 \\
     SN2003dh & 1.5 & 0.09 & 61.14  & 27.05\\
     SN2007bi & 11.9 & 4.18 & 64.07 & 63.80\\
     SN2015U & 1.3 & 0.44 & 1.47 & 1.13\\
     SN2016coi & 4.8 & 3.61  & 14.00 & 13.11 \\
     SN2019dge & 0.7 & 0.17  & 2.89  & 2.98 \\
    \hline      
\end{tabular}
\end{table}

\begin{table}
\centering
\caption{Ejected masses of SESNe calculated from the photospheric velocities ($v_{ph}$)}
\label{tab:mass2}
\begin{tabular}{lcccc}
\hline
  SN Name & $M_{ej}$ ($M_\odot$)  & $M_{peak}$ ($M_\odot$) & $M_\gamma$ ($M_\odot$) & $M_{+}$ ($M_\odot$) \\
 \hline
     SN1993J & 2.1 & 1.23 & 4.16 & 4.08\\
     SN1996cb & 2.8 & 1.48 &  5.49 & 5.22\\
     SN2006T & 2.3 & 1.14 & 3.26 & 3.10 \\
     SN2006el & 3.8 & 1.36 & 9.38 & 9.36\\
     SN2008ax & 1.77 & 0.78 & 3.33 & 3.09\\
     SN2009mg & 1.9  & 1.72 & 8.30 & 7.53\\
     SN2011dh & 1.4  & 0.71 & 3.38 & 3.19\\
     SN2011fu & 1.0  & 0.95 & 24.78 & 24.68 \\
     SN2011hs & 2.6  & 1.07 & -- & -- \\
     SN2013df & 2.37 & 1.70 & 1.63 & 2.91 \\
    \hline 
     SN1983N & 2.8 & 2.06 & 7.59 & 5.87 \\
     SN1999dn & 3.5  & 1.52  & 35.09  & 35.09 \\
     SN2004gq & 3.4  & 1.72  & 15.35 & 12.33\\
     SN2007C & 2.3 & 0.82 & 8.80 & 8.00 \\
     SN2007uy & 6.2 & 13.43  & 27.44  & 26.40 \\
     SN2008D & 3.4 & 2.89  & 5.16 & 5.14 \\
     SN2009jf & 4.8 & 7.91 & 11.16 & 10.82 \\
     PTF13bvn & 4.8 & 4.46 & 7.85 & 7.79\\
    \hline
     SN1962L & 2.9 & 0.70 & 10.44  & 0.41 \\
     SN1998bw & 4.7 & 7.13 & 26.99 & 25.36\\
     SN2002ap & 6.8 & 3.92  & 19.93 & 19.87 \\
     SN2005hg & 3.0  & 2.56 & -- & --\\
     SN2006ep & 3.8  & 2.11 & 9.02 & 8.48\\
     SN2007Y & 2.94  & 1.66  & 2.95 & 2.01\\
     SN2007gr & 4.7  & 2.31  & 4.25 & 3.98\\
    \hline
     SN1994I & 2.8 & 1.46  & 6.39 & 5.87 \\
     SN2004aw & 6.5 & 10.60 & 187.73 & 176.18\\
     SN2004dn & 2.0  & 1.59 & 10.10 & 9.43 \\
     SN2004fe & 2.3  & 1.65  & 1.41  & 1.14 \\
     SN2004ff & 2.8  & 1.07  & -- & -- \\
     SN2009bb & 2.7  & 2.92  & 6.16  & 5.32 \\
     SN2010bh & 2.1  & 3.94  & -- & --\\
     SN2010gx & 4.75  & 1.28 & 4.25 & 2.37 \\
     SN2011bm & 6.6  & 5.35  & 3.10 & 2.54\\
     SN2013dg & 3.92  & 0.14  & 36.53  & 13.21 \\
     SN2013ge & 3.2  & 1.06  & 19.24 & 19.17 \\
     SN2014L & 2.25  & 0.42  & -- & --\\
     SN2015bn & 13.8  & 4.57  & --  & 18.06 \\
     iPTF15dtg & 4.2  & 2.23  & -- & --\\
     SN2017ein & 2.2  & 1.09  & -- & --\\
    \hline
     SN1997ef & 3.8 & 2.06 & 6.14 & 5.83 \\
     SN2016coi & 4.8 & 5.05 & 27.44  & 25.69\\
     SN2019dge & 0.7 & 0.15 & 2.28  & 2.35 \\
    \hline      
\end{tabular}
\end{table}

\begin{table}
\centering
\caption{Ejected masses of SESNe calculated from the scaling velocities ($v_{sc}$)}
\label{tab:mass3}
\begin{tabular}{lcccc}
\hline
  SN Name & $M_{ej}$ ($M_\odot$)  & $M_{peak}$ ($M_\odot$) & $M_\gamma$ ($M_\odot$) & $M_{+}$ ($M_\odot$) \\
 \hline
     SN1993J & 2.1 & 2.32 & 14.69 & 14.41\\
     SN1996cb & 2.8 & 1.80 & 8.06 & 7.65\\
     SN2004ex & 2.6 & 2.54 & 10.29 & 9.28\\
     SN2006T & 2.3 & 2.66& 17.75 & 16.89\\
     SN2006el & 3.8 & 1.59 & 12.90 & 12.88\\
     SN2008aq & 2.6 & 1.26 & 9.71 & 9.27\\
     SN2008ax & 1.77 & 2.44  & 32.80 & 30.40\\
     SN2009mg & 1.9 & 2.76 & 21.25  & 19.27\\
     SN2011dh & 1.4  & 2.23 & 33.77 & 31.82\\
     SN2011fu & 1.0 & 1.41  & 55.09  & 54.86\\
     SN2011hs & 2.6 & 1.88 & -- & --\\
     SN2012P & 1.85 & 1.27 & 28.31  & 21.22 \\
     SN2013df & 2.37 & 2.17 & 2.66  & 4.76\\
     SN2016gkg & 3.6 & 2.63 & 33.42 & 33.29\\ 
    \hline 
     SN1983N & 2.8  & 2.74 & 13.50 & 10.44\\
     SN1999dn & 3.5 & 1.76 & 47.22  & 47.22\\
     SN2004gq & 3.4 & 0.93  & 4.45  & 3.58\\
     SN2007C & 2.3  & 1.08 & 15.29 & 13.90\\
     SN2007uy & 6.2  & 4.51 & 3.09 & 2.98\\
     SN2008D & 3.4 & 2.48  & 3.82 & 3.80\\
     SN2009jf & 4.8 & 4.67 & 3.89 & 3.77\\
     PTF13bvn & 4.8 & 3.85 & 5.83 & 5.79\\
    \hline
     SN1962L & 2.9 & 1.15  & 28.25 & 1.12\\
     SN1998bw & 4.7  & 4.18 & 9.26  & 8.70\\
     SN1999ex & 3.0 & 1.84 & 2.14 & 1.55\\
     SN2002ap & 6.8  & 2.54 & 8.33  & 8.31\\
     SN2005hg & 3.0  & 2.70 & -- & --\\
     SN2006ep & 3.8 & 1.98  & 7.91 & 7.45\\
     SN2007Y & 2.94 & 2.28 & 5.59 & 3.81\\
     SN2007gr & 4.7 & 2.17 & 3.73 & 3.50\\
     SN2009iz & 4.1 & 2.30& 3.39  & 3.38\\
     DES17C1ffz & 3.7 & 4.89 & 9.33 & 9.21\\
    \hline
     SN1994I & 2.8  & 1.03 & 3.17 & 2.91\\
     SN2004aw & 6.5  & 5.63 & 52.98  & 49.72\\
     SN2004dn & 2.0   & 1.53 & 9.31 & 8.69\\
     SN2004fe & 2.3 & 2.15  & 2.39  & 1.93\\
     SN2004ff & 2.8  & 0.82 & -- & -- \\
     SN2006lc & 2.91  & 1.20 & 13.84 & 11.44\\
     SN2007cl & 2.99 & 1.34 & 2.98 & 2.77\\
     SN2009bb & 2.7  & 1.77 & 2.26 & 1.95\\
     SN2010bh & 2.1  & 1.52  & -- & --\\
     SN2010gx & 4.75 & 0.45  & 0.52  & 0.29\\
     SN2010md & 6.4  & 6.24 & 11.46 & 7.35\\
     SN2011bm & 6.6  & 6.41  & 4.46 & 3.66\\
     SN2011ke & 5.42 & 1.15 & 0.62 & 0.53\\
     SN2011kg & 4.5  & 1.27 & -- &--\\
     SN2013dg & 3.92  & 0.06 & 8.15 & 2.94\\
     SN2013ge & 3.2  & 1.27 & 27.70 & 27.61\\
     SN2014L & 2.25  & 0.81 & -- & --\\
     SN2014ad & 2.6 & 0.70 & 9.67 & 9.23 \\
     SN2015bn & 13.8  & 6.89 & -- & 41.05\\
     iPTF15dtg & 4.2 & 4.02 & -- & --\\
     SN2017ein & 2.2 & 1.74 & -- & --\\
    \hline
     SN1997ef & 3.8 & 1.99 & 5.75  & 5.47\\
     SN2003dh & 1.5  & 0.10  & 75.34 & 33.33\\
     SN2007bi & 11.9 & 2.09 & 16.02 & 15.95\\
     SN2015U & 1.3 & 0.90 & 6.29 & 4.84\\
     SN2016coi & 4.8 & 2.35  & 5.91 & 5.54\\
     SN2019dge & 0.7  & 4.83 & 23.01 & 23.71\\
    \hline      
\end{tabular}
\end{table}

\begin{figure}
	\includegraphics[width=\columnwidth]{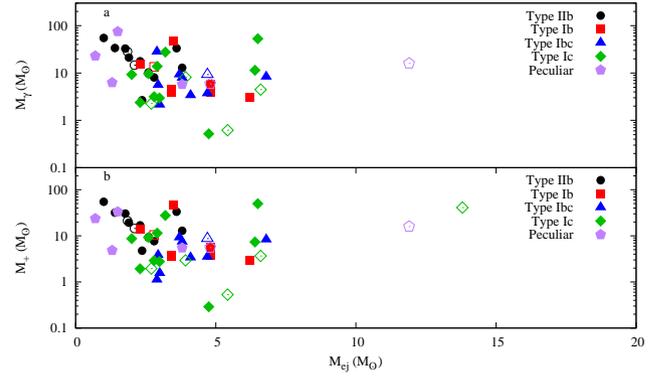}
    \caption{Comparison of $M_{ej}$ with $M_{\gamma}$ (panel a) and $M_+$ (panel b) derived from scaling velocities. The different colors represent the different supernovae types (IIb black, Ib red, Ibc blue, Ic green and peculiar purple), while the empty and filled symbols show group 1 and group 2 SNe within a particular type.}
    \label{fig:mass}
\end{figure}

Overestimation at late-times could be an effect of an unreliable velocity approximation or a misleading LC fit. Both scenarios suggest an interaction with CSM, which may change the slope of the light curve (see group 1 SNe fits) and also reduce the expansion velocity. Another indication of this process can be seen in Fig. \ref{fig:mass} which demonstrates a comparison between the derived ejecta masses from scaling velocities. Here, group 1 (objects with irregularities in their LCs) and group 2 SNe do not show any difference, which suggests that all stripped-envelope supernovae, regardless of their visible LCs features, are involved in CSM interaction. This speculation also can be supported by the single-star progenitor theory, where we assume an extreme mass-loss at the late evolutionary phase.

If we assume a CSM around the exploding star, the generally used homologous expansion may be inaccurate at late times which indicates that we should scale down the velocities for these calculations. However, taking into account the velocity changes could be problematic because we are not able to determine the physical properties of the CSM from our model.

According to basic analytical studies, during the interaction phase, the ratio of late-time and initial velocities ($\alpha$) is about 0.5 - 0.8 depending on the CSM density profile, while in the Sedov phase $\alpha \sim$ 0.4. The time-scale of these phases could be really different due to the intensity of the CSM interaction. To test our results we systematically calculated this parameter using the assumption that $\alpha \approx v_{mod}/v_{sc} = \sqrt{M_+/M_{ej}}$. We also exclude events with extremely short or uncertain late-time light curves, where $T_0$ values could be inconsistent. Then, we were able to compare the derived parameters and the theoretical predictions (Table \ref{tab:v_mod}). Then, we presume that the average $\alpha$ for a particular supernova type could be a reasonable velocity modification constant. For Type IIb, Ib, Ibc, Ic and peculiar SNe this value is 0.39, 0.74, 0.82, 0.67 and 0.87, respectively. Using these we can recalculate the scaling velocities and ejected masses ($M_{+,mod}$) that are now closer to the expected $M_{ej}$ values in most cases. Thus, with this empirical scaling method, we may reduce the tension between the early and late-time mass estimations.

\begin{table}
\centering
\caption{Modified scaling velocities and ejected masses of SESNe}
\label{tab:v_mod}
\begin{tabular}{lcccc}
\hline
  SN Name & $\alpha$ & $v_{mod}$ ($km/s$) & $M_{+,mod}$ ($M_\odot$) & $M_{ej}$ ($M_\odot$)\\
 \hline
     SN1993J & 0.38 & 5499 & 2.19  & 2.1 \\ 
     SN1996cb & 0.60 & 4251 & 1.16 & 2.8 \\
     SN2006T & 0.37 & 6825 & 2.57 & 2.3 \\
     SN2006el & 0.54 & 5031 & 1.96 & 3.8 \\
     SN2008aq & 0.53 & 5343 & 1.41 & 2.6 \\
     SN2008ax & 0.24 & 9789 & 4.62 & 1.77\\
     SN2011dh & 0.21 & 8502 & 4.84 & 1.4\\
     SN2012P & 0.30 & 6318 & 3.23 & 1.85  \\
     SN2016gkg & 0.33 & 4680 & 5.06 & 3.6\\ 
    \hline 
     SN1983N & 0.52 & 7696 & 5.72 & 2.8 \\
     SN1999dn & 0.27 & 8584 & 25.9 & 3.5 \\
     SN2004gq & 0.97 & 5180  & 1.96 & 3.4 \\
     SN2007C & 0.41 & 10730  & 7.61 & 2.3 \\
     SN2008D & 0.95 & 6364 & 2.08 & 3.4 \\
     SN2009jf & 1.13 & 4366  & 2.06 & 4.8 \\
     PTF13bvn & 0.91 & 5550 & 3.17  & 4.8 \\
    \hline
     SN1998bw & 0.74 & 6724 & 5.85 & 4.7\\
     SN2002ap & 0.90 & 7954  & 5.59 & 6.8 \\
     SN2006ep & 0.71 & 7298  & 5.01 & 3.8 \\
     SN2007Y & 0.88 & 10168 & 2.56 & 2.94\\
     SN2007gr & 1.16 & 4920  & 2.35 & 4.7 \\
     DES17C1ffz & 0.63 & 9430 & 6.19 & 3.7\\
    \hline
     SN2004dn & 0.48 & 8040 & 3.9 & 2.0 \\
     SN2010md & 0.93 & 5226 & 3.3 & 6.4 \\
     SN2013dg & 1.15 & 5695 & 1.32 & 3.92 \\
     SN2013ge & 0.34 & 8040 & 12.39 & 3.2\\
     SN2014ad & 0.53 & 9179 & 4.14 & 2.6 \\
     SN2015bn & 0.58 & 6566  & 18.4 & 13.8 \\
    \hline
     SN1997ef & 0.83 & 8004  & 4.14 & 3.8 \\
     SN2007bi & 0.86 & 4350 & 12.1 & 11.9\\
     SN2016coi & 0.93 & 5655 & 4.19 & 4.8 \\
    \hline      
\end{tabular}
\end{table}

\section{Conclusions}
We have presented the results of our systematic study related to stripped-envelope supernova modeling. Our main goal was testing the generally used model approximations in order to make sure that the mass-discrepancy problem is not caused by the initial condition of the applied analytical models. 

First, we tested the R-band approximation and our study revealed that in most cases the R-band and the bolometric light curve do not show the same general features (e.g. rise time, decline rate of the tail). Thus, this approximation is not sufficient to estimate
the physical properties of individual supernovae. However, the mass ratio of early- and late-time ejected mass calculations do not show significant differences, which indicates that this approximation does not cause the mass-discrepancy problem.

We also took into consideration the rise-time approximation. The main problem with this method is that, except for some recent SESNe, the explosion times are quite uncertain, which causes the underestimation of the ejected mass. However, due to recent all-sky surveys, the determination of the supernova explosion date could be more precise in the near future. Till then, fitting the early light curve and defining the ejected mass from it seems to be a bit more reasonable solution. But, there are some issues in our model to fit the rising phase of the supernova LCs. The solution for this could be an alternative energy source or a two-component model, as our previous study revealed  \citep[e.g.,][]{Nagy1} for Type IIb supernovae.

The third estimation we tested is related to the ejected mass calculation of late-time supernova light curves. The results show that taking into account both gamma-ray and positron-leakage somewhat reduces the derived ejecta masses due to the lower $T_0$ values as well as giving more reasonable nickel masses. But, this scenario still overestimates $M_{ej}$ regardless of the used velocity definition. To moderate this issue we may modify the velocity and/or opacity values or implement a self-consistent $T_0$ calculation approach in our code.

Overall, our results suggest that the mass-discrepancy problem is not strongly related to the basic approximations, but the mass difference can be reduced with fine-tuning of these estimations, as we show above. Moreover, this also indicates that we should study the physical structure of these objects in more detail and improve our model to consider these characteristics as accurately as possible. 

\newpage
\section*{Acknowledgements}
Special thanks to John Craig Wheeler for the helpful suggestions and discussions related to the mass-discrepancy problem and this article.
This project is supported by NKFIH/OTKA PD- 134434 grant, which is founded by the Hungarian National Development and Innovation Fund. 

\section*{Data Availability}

The data underlying this article are available in the Open Supernova Catalog, currently at  https://github.com/astrocatalogs.
While, the source code of the program applied for light curve fitting in this paper is available online at http://titan.physx.u-szeged.hu/~nagyandi/LC2.




\bibliographystyle{mnras}
\bibliography{example} 

\begin{thebibliography}{}
\makeatletter
\relax
\def\mn@urlcharsother{\let\do\@makeother \do\$\do\&\do\#\do\^\do\_\do\%\do\~}
\def\mn@doi{\begingroup\mn@urlcharsother \@ifnextchar [ {\mn@doi@}
  {\mn@doi@[]}}
\def\mn@doi@[#1]#2{\def\@tempa{#1}\ifx\@tempa\@empty \href
  {http://dx.doi.org/#2} {doi:#2}\else \href {http://dx.doi.org/#2} {#1}\fi
  \endgroup}
\def\mn@eprint#1#2{\mn@eprint@#1:#2::\@nil}
\def\mn@eprint@arXiv#1{\href {http://arxiv.org/abs/#1} {{\tt arXiv:#1}}}
\def\mn@eprint@dblp#1{\href {http://dblp.uni-trier.de/rec/bibtex/#1.xml}
  {dblp:#1}}
\def\mn@eprint@#1:#2:#3:#4\@nil{\def\@tempa {#1}\def\@tempb {#2}\def\@tempc
  {#3}\ifx \@tempc \@empty \let \@tempc \@tempb \let \@tempb \@tempa \fi \ifx
  \@tempb \@empty \def\@tempb {arXiv}\fi \@ifundefined
  {mn@eprint@\@tempb}{\@tempb:\@tempc}{\expandafter \expandafter \csname
  mn@eprint@\@tempb\endcsname \expandafter{\@tempc}}}

\bibitem[\protect\citeauthoryear{Arnett \& Fu}{Arnett \& Fu}{1989}]{Arnett}
Arnett W.~D.,  Fu A.,  1989, ApJ, 340, 396

\bibitem[\protect\citeauthoryear{Bessell, Castelli  \& Plez}{Bessell
  et~al.}{1998}]{Bessell}
Bessell M.~S.,  Castelli F.,   Plez B.,  1998, A\&A, 333, 231

\bibitem[\protect\citeauthoryear{Clocchiatti \& Wheeler}{Clocchiatti \&
  Wheeler}{1997}]{Clocchiatti}
Clocchiatti A.,  Wheeler J.~C.,  1997, ApJ, 491, 375

\bibitem[\protect\citeauthoryear{{Colgate}, {Petschek}  \& {Kriese}}{{Colgate}
  et~al.}{1980}]{Colgate}
{Colgate} S.,  {Petschek} A.,   {Kriese} J.,  1980, ApJL, 237, 81

\bibitem[\protect\citeauthoryear{Dessart, Hillier, Livne, Yoon, Woosley,
  Waldman  \& Langer}{Dessart et~al.}{2011}]{Dessart}
Dessart L.,  Hillier D.~J.,  Livne E.,  Yoon S.-C.,  Woosley S.,  Waldman R.,
  Langer N.,  2011, MNRAS, 414, 2985–3005

\bibitem[\protect\citeauthoryear{Eldridge, Izzard  \& Tout}{Eldridge
  et~al.}{2008}]{Eldridge}
Eldridge J.~J.,  Izzard R.~G.,   Tout C.~A.,  2008, MNRAS, 384, 1109

\bibitem[\protect\citeauthoryear{Filippenko}{Filippenko}{1997}]{Filippenko}
Filippenko A.~V.,  1997, ARA\&A, 35, 309

\bibitem[\protect\citeauthoryear{Gal-Yam}{Gal-Yam}{2017}]{Gal-Yam}
Gal-Yam A.,  2017, https://doi.org/10.1007/978-3-319-21846-5\_35, p.~195

\bibitem[\protect\citeauthoryear{Guillochon, Parrent, Kelley  \&
  Margutti}{Guillochon et~al.}{2017}]{Guillochon}
Guillochon J.,  Parrent J.,  Kelley L.~Z.,   Margutti R.,  2017, ApJ, 835, 64

\bibitem[\protect\citeauthoryear{Inserra et~al.,}{Inserra
  et~al.}{2013}]{Inserra}
Inserra C.,  et~al., 2013, ApJ, 770, 128

\bibitem[\protect\citeauthoryear{Kasen \& Bildsten}{Kasen \&
  Bildsten}{2010}]{Kasen}
Kasen D.,  Bildsten L.,  2010, ApJ, 717, 245

\bibitem[\protect\citeauthoryear{Khatami \& Kasen}{Khatami \&
  Kasen}{2019}]{Khatami}
Khatami D.~K.,  Kasen D.~N.,  2019, ApJ, 878, 56

\bibitem[\protect\citeauthoryear{Lyman, Bersier  \& James}{Lyman
  et~al.}{2014}]{Lyman14}
Lyman J.~D.,  Bersier D.,   James P.~A.,  2014, MNRAS, 437, 3848

\bibitem[\protect\citeauthoryear{Lyman, Bersier, James, Mazzali, Eldridge,
  Fraser  \& Pian}{Lyman et~al.}{2016}]{Lyman}
Lyman J.~D.,  Bersier D.,  James P.~A.,  Mazzali P.~A.,  Eldridge J.~J.,
  Fraser M.,   Pian E.,  2016, MNRAS, 457, 328–350

\bibitem[\protect\citeauthoryear{Nadyozhin}{Nadyozhin}{1994}]{Nadyozhin}
Nadyozhin D.~K.,  1994, ApJSS, 92, 527

\bibitem[\protect\citeauthoryear{Nagy}{Nagy}{2018}]{Nagy2}
Nagy A.~P.,  2018, ApJ, 862, 143

\bibitem[\protect\citeauthoryear{Nagy \& Vinkó}{Nagy \& Vinkó}{2016}]{Nagy1}
Nagy A.~P.,  Vinkó J.,  2016, A\&A, 589, A53

\bibitem[\protect\citeauthoryear{Nagy, Ordasi, Vinkó  \& Wheeler}{Nagy
  et~al.}{2014}]{Nagy}
Nagy A.~P.,  Ordasi A.,  Vinkó J.,   Wheeler J.~C.,  2014, A\&A, 571, A77

\bibitem[\protect\citeauthoryear{Podsiadlowski, Joss  \& Hsu}{Podsiadlowski
  et~al.}{1992}]{Podsiadlowski}
Podsiadlowski P.,  Joss P.~C.,   Hsu J. J.~L.,  1992, ApJ, 391, 246

\bibitem[\protect\citeauthoryear{Prentice et~al.,}{Prentice
  et~al.}{2016}]{Prentice}
Prentice S.~J.,  et~al., 2016, MNRAS, 458, 2973–3002

\bibitem[\protect\citeauthoryear{{Sawada} \& {Maeda}}{{Sawada} \&
  {Maeda}}{2019}]{Sawada}
{Sawada} R.,  {Maeda} K.,  2019, \apj, 886, 47

\bibitem[\protect\citeauthoryear{Shivvers et~al.,}{Shivvers
  et~al.}{2017}]{Shivvers}
Shivvers I.,  et~al., 2017, PASP, 129, 054201

\bibitem[\protect\citeauthoryear{Smartt, Eldridge, Crockett  \& Mound}{Smartt
  et~al.}{2009}]{Smartt}
Smartt S.~J.,  Eldridge J.~J.,  Crockett R.~M.,   Mound J.~R.,  2009, MNRAS,
  395, 1409

\bibitem[\protect\citeauthoryear{Sollerman et~al.,}{Sollerman
  et~al.}{2020}]{Sollerman}
Sollerman J.,  et~al., 2020, A\&A, 643, A79

\bibitem[\protect\citeauthoryear{Taddia et~al.,}{Taddia et~al.}{2015}]{Taddia}
Taddia F.,  et~al., 2015, A\&A, 547, A60

\bibitem[\protect\citeauthoryear{Turatto \& Pastorello}{Turatto \&
  Pastorello}{2013}]{Turatto}
Turatto M.,  Pastorello A.,  2013, Proceedings of the International
  Astronomical Union, 9, 63

\bibitem[\protect\citeauthoryear{Wheeler, Johnson  \& Clocchiatti}{Wheeler
  et~al.}{2015}]{Wheeler}
Wheeler J.~C.,  Johnson V.,   Clocchiatti A.,  2015, MNRAS, 450, 1295

\makeatother
\end{thebibliography}




\appendix 
\section{Light curve fits of SESNe}

Here, we show the bolometric light curve fits of all examined SESNe. In each figure, the blue line represents the best model for the entire LC due to the fit-by-eye method, while the green and black lines show the calculation results for the two different late-time fitting approaches.

 \begin{figure*}
	\includegraphics[]{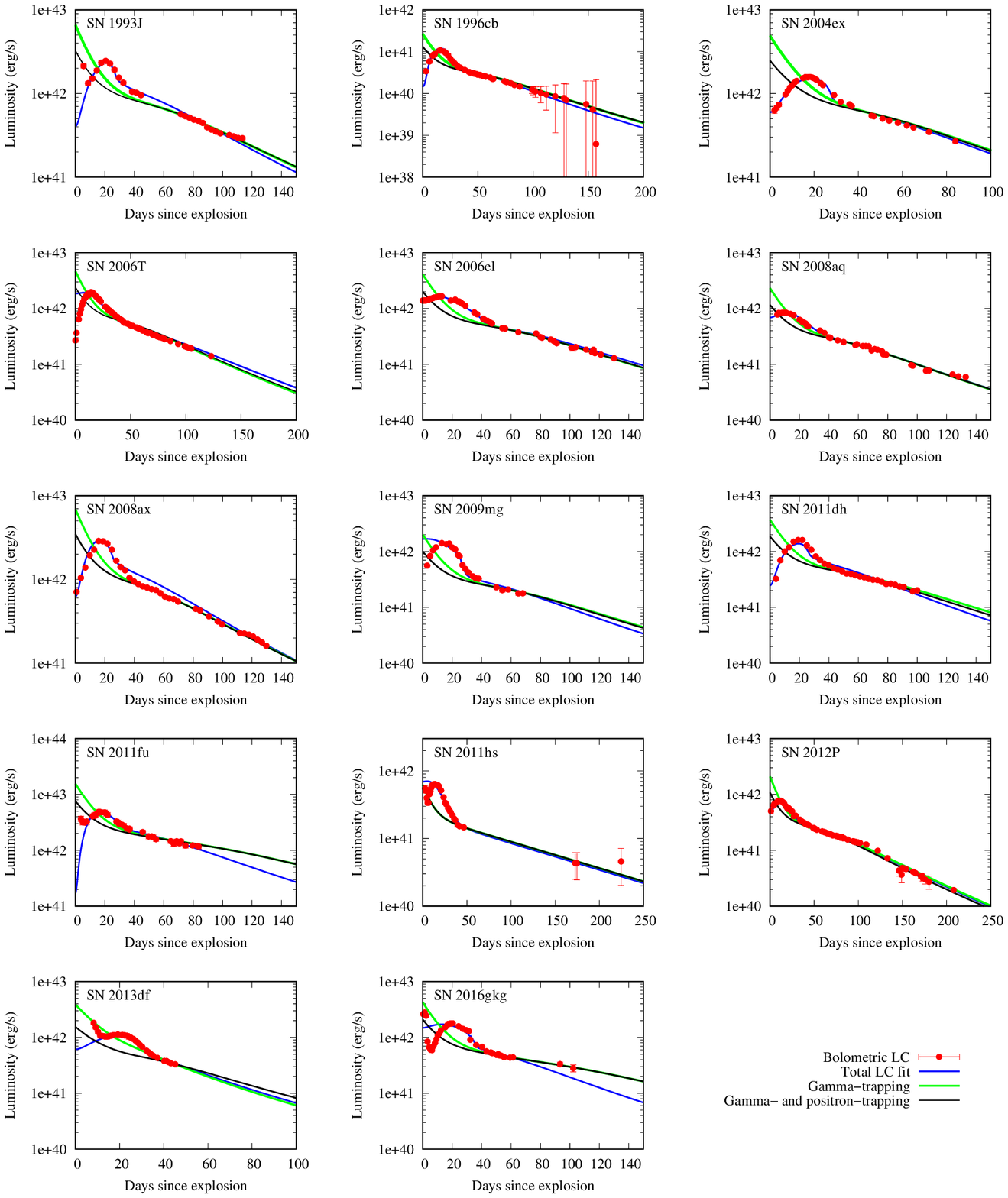}
    \caption{Bolometric light curve fits for Type IIb supernovae. The red dots show the observational data with their photometric errors. The blue line represents the best fit for the entire light curve, while the green and black lines show the late-time LC fits for only gamma-leakage and gamma- and positron-leakage, respectively.}
    \label{fig:lc1}
\end{figure*}

 \begin{figure*}
	\includegraphics[]{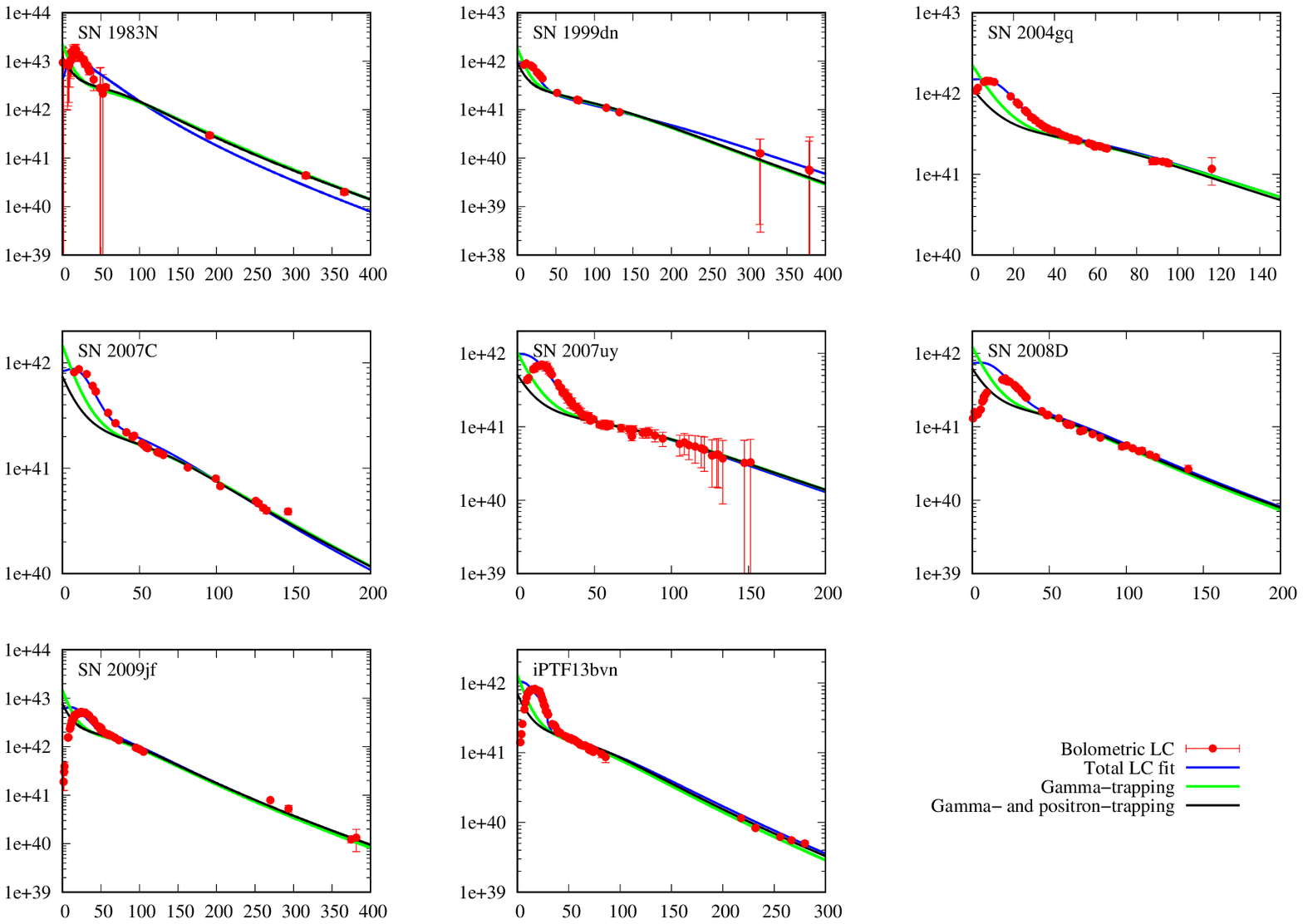}
    \caption{Bolometric light curve fits for Type Ib supernovae. The red dots show the observational data with their photometric errors. The blue line represents the best fit for the entire bolometric light curve, while the green and black lines show the late-time LC fits for only gamma-leakage and gamma- and positron-leakage, respectively.}
    \label{fig:lc2}
\end{figure*}

 \begin{figure*}
	\includegraphics[]{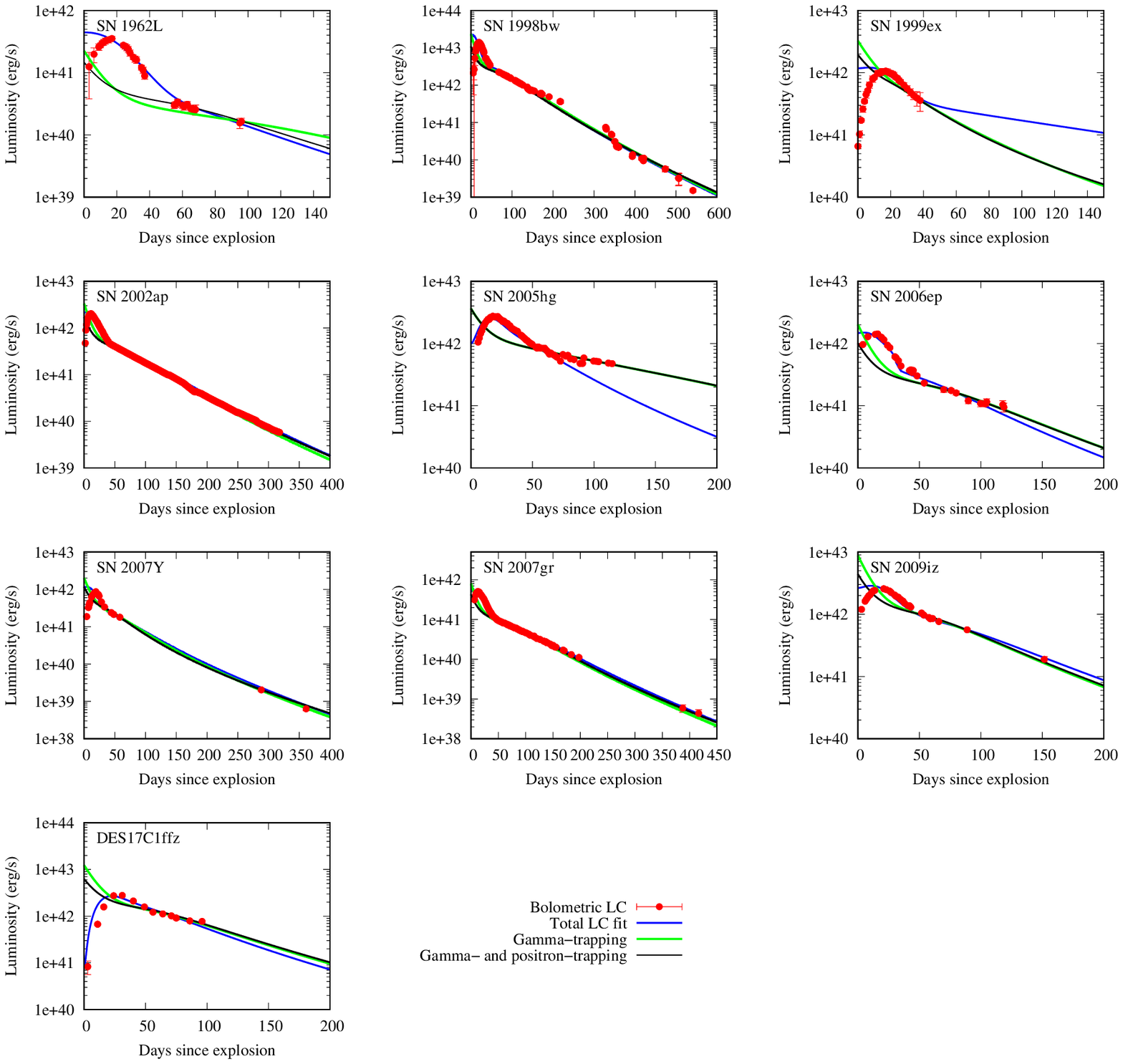}
    \caption{Bolometric light curve fits for Type Ibc supernovae. The red dots show the observational data with their photometric errors. The blue line represents the best fit for the entire bolometric light curve, while the green and black lines show the late-time LC fits for only gamma-leakage and gamma- and positron-leakage, respectively.}
    \label{fig:lc3}
\end{figure*}

 \begin{figure*}
	\includegraphics[]{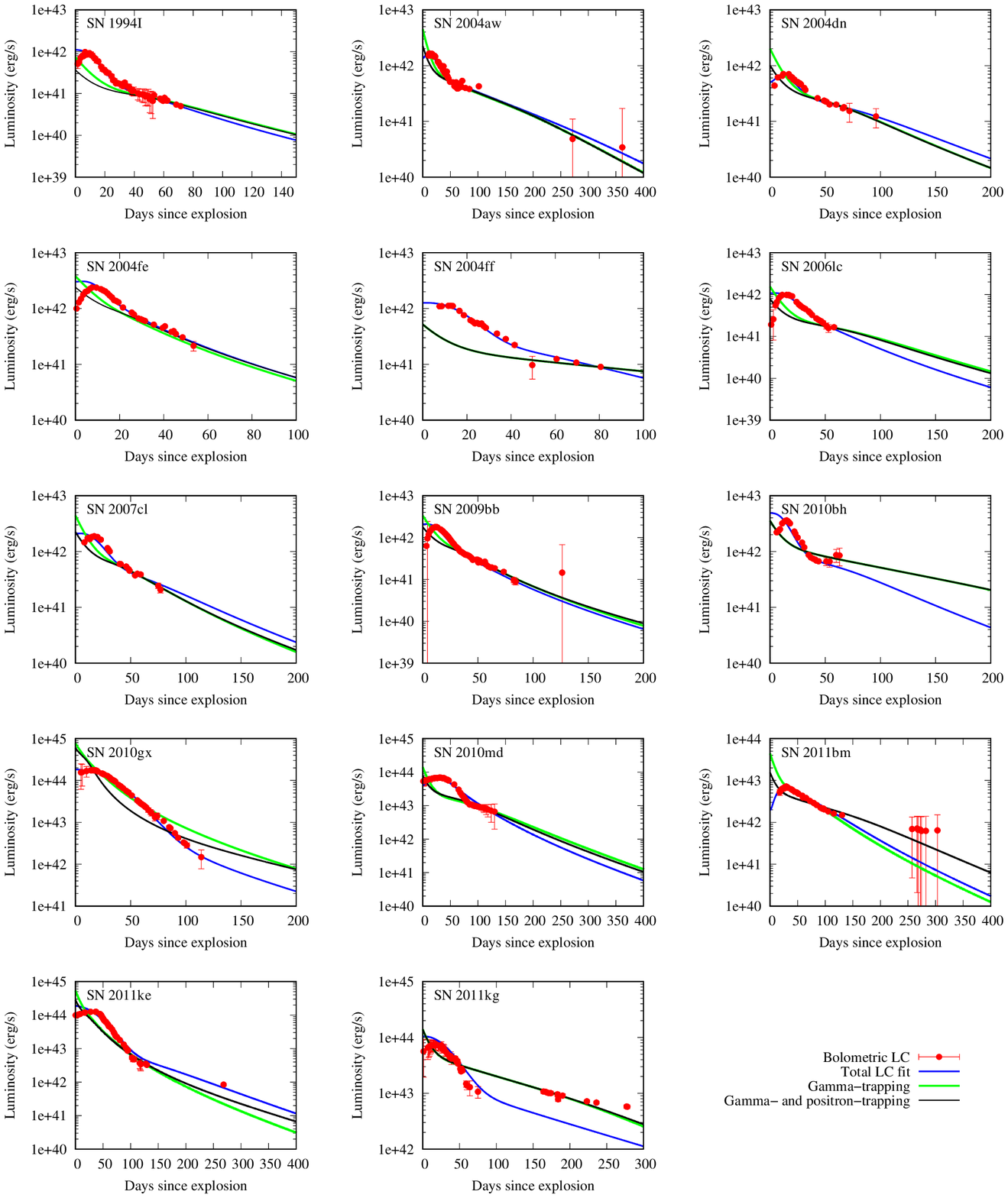}
    \caption{Bolometric light curve fits for Type Ic supernovae. The red dots show the observational data with their photometric errors. The blue line represents the best fit for the entire bolometric light curve, while the green and black lines show the late-time LC fits for only gamma-leakage and gamma- and positron-leakage, respectively.}
    \label{fig:lc4}
\end{figure*}

 \begin{figure*}
	\includegraphics[]{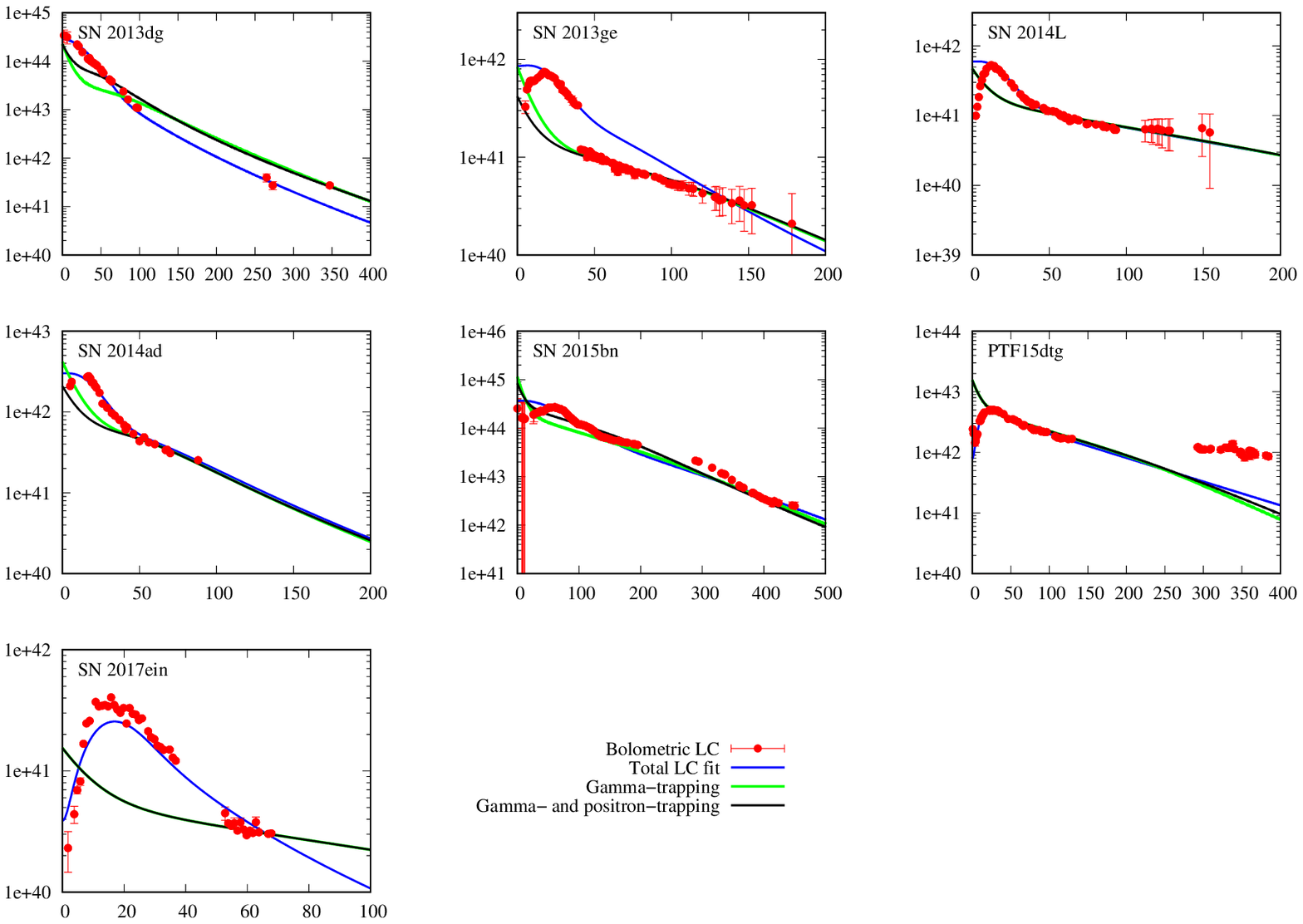}
    \caption{Bolometric light curve fits for Type Ic supernovae. The red dots show the observational data with their photometric errors. The blue line represents the best fit for the entire bolometric light curve, while the green and black lines show the late-time LC fits for only gamma-leakage and gamma- and positron-leakage, respectively.}
    \label{fig:lc5}
\end{figure*}

 \begin{figure*}
	\includegraphics[]{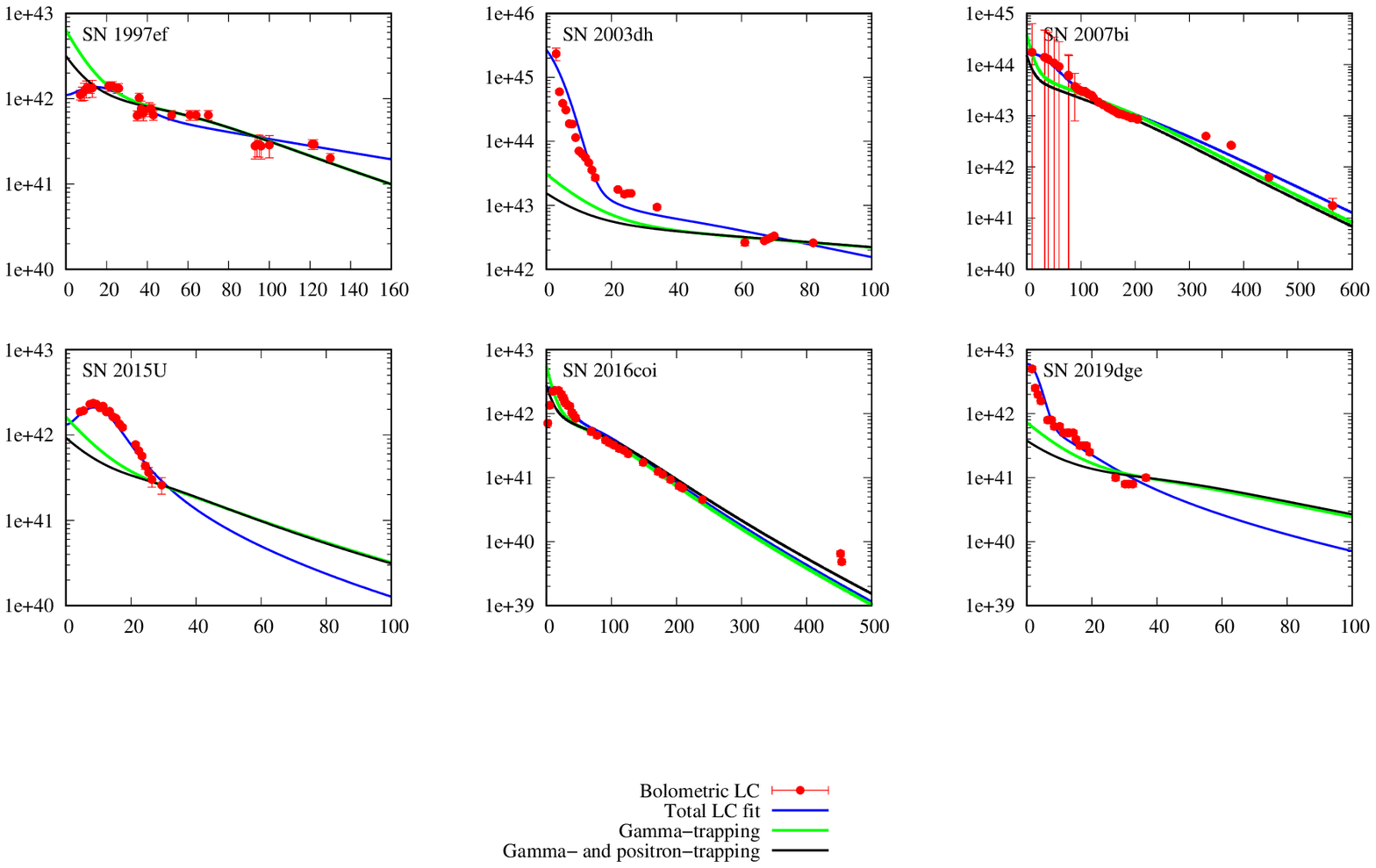}
    \caption{Bolometric light curve fits for peculiar SESNe. The red dots show the observational data with their photometric errors. The blue line represents the best fit for the entire bolometric light curve, while the green and black lines show the late-time LC fits for only gamma-leakage and gamma- and positron-leakage, respectively.}
    \label{fig:lc6}
\end{figure*}


\bsp	
\label{lastpage}
\end{document}